\documentclass[lettersize,journal]{IEEEtran}
\usepackage{cite}
\usepackage{xcolor}
\usepackage[switch,columnwise]{lineno}
\usepackage{amsmath,amsfonts}
\usepackage{array}
\usepackage[caption=false,font=normalsize,labelfont=sf,textfont=sf]{subfig}
\usepackage{textcomp}
\usepackage{stfloats}
\usepackage{verbatim}
\usepackage{tikz}
\usepackage{amsmath}
\usepackage{pgfplots}
\pgfplotsset{compat=1.18}
\usepackage{longtable}
\usepackage{booktabs}
\usepackage{amssymb}
\usepackage{pifont}
\usepackage{graphicx}
\usepackage{microtype}
\usepackage{algorithm}
\usepackage{algpseudocode}
\usepackage{xtab}
\usepackage{xurl}
\usepackage{makecell}
\usepackage{multirow}
\microtypecontext{spacing=nonfrench}
\usepackage{listings}
\usepackage{float}
\newfloat{listing}{t}{lop}
\floatname{listing}{Listing}
\definecolor{figblue}{RGB}{0, 0, 255} 
\definecolor{figgreen}{RGB}{50, 168, 82}

\begin{document}

\title{Beyond the Monolithic Scheduler: Lightweight Adaptive Policy Switcher for Process Scheduling Optimization}

\author{Xinbo Wang,
        Shian Jia,
        Ziyang Huang,
        Jing Cao,
        and Mingli Song%
\thanks{This work is supported in part by the Hangzhou Joint Funds of the Zhejiang Provincial Natural Science Foundation of China under Grant No. LHZSD24F020001, in part by the Zhejiang Provincial Natural Science Foundation of China under Grant No. LMS26F020046, in part by the Fundamental Research Funds for the Central Universities under Grant No. 226-2025-00067, and in part by the advanced computing resources provided by the Supercomputing Center of Hangzhou City University. \textit{(Corresponding author: Jing Cao.)} Xinbo Wang and Shian Jia contributed equally to this research.}%
\thanks{Xinbo Wang, Shian Jia, Ziyang Huang, and Mingli Song are with Zhejiang University, Hangzhou 310027, China (e-mail: xinbowang@zju.edu.cn; csjsa@zju.edu.cn; zyanghuang@zju.edu.cn; brooksong@zju.edu.cn).}%
\thanks{Jing Cao is with the High-Performance Intelligent Computing Research Center for Ultra-Large Scale Graph Data, School of Computer and Computing Science, Hangzhou City University, Hangzhou 310015, China, and also with the State Key Laboratory of Blockchain and Data Security, Zhejiang University, Hangzhou 310027, China (e-mail: cjcongjia@zju.edu.cn).}%
}
\markboth{IEEE TRANSACTIONS ON COMPUTER-AIDED DESIGN OF INTEGRATED CIRCUITS AND SYSTEMS}{%
  Wang \MakeLowercase{\textit{et al.}}: Beyond the Monolithic Scheduler}
\maketitle

\begin{abstract}
Modern operating system process schedulers typically enforce a static, ``one-policy-fits-all'' design, which inevitably compromises performance under the highly dynamic and mixed workloads of contemporary edge systems. Transitioning to dynamic adaptation faces two fundamental challenges: scenario-specific static policies lack robust runtime adaptability, whereas computationally intensive AI-assisted policies introduce unacceptable execution overhead for resource-constrained devices. To structurally resolve this dilemma, we propose the Lightweight Adaptive Policy Switcher (LAPS), a framework that dynamically switches between predefined expert policies, thus decoupling high-frequency context switching from low-frequency policy transitions and amortizing transitional overheads. 
To eliminate prohibitive online computations, LAPS pre-resolves all heavy dynamic trade-offs via an offline preparation phase, constructing a strictly overhead-aware Policy Mapping Table. During runtime, it executes an agentic Observe-Infer-Switch cycle that limits machine learning inference strictly to workload classification. Consequently, LAPS executes efficient dynamic adaptation simply by mapping the inferred workload state to optimal expert policies via low-cost table lookups.
Experiments show that LAPS achieves an average performance gain of 8.91\% and reduces response latency by up to 91.7\% in latency-sensitive tasks over the default Linux scheduler EEVDF. 
Our results demonstrate that LAPS outperforms the baseline in a wide range of test cases, offering a practical path toward more responsive and intelligent operating systems. Our code is publicly available at \url{https://github.com/lvmiao233/LAPS}.
\end{abstract}


\section{Introduction}
Operating system process scheduling functions as the critical last mile of resource arbitration. Unlike other resource scheduling layers, such as cluster management or Virtual Machine (VM) orchestration, which typically operate on \emph{seconds-level} timescales with visibility into application semantics, the kernel scheduler operates strictly in the \emph{microsecond} domain with limited high-level context. It must arbitrate access to shared silicon resources (e.g., ALU, L1/L2 caches) for individual instruction streams without the capacity for lengthy decision cycles. Due to this constraint of making high-frequency decisions based on low-level signals, scheduler development has historically focused on refining heuristics to establish a generalized balance among fairness, latency, and throughput.

However, the complexity of modern intelligent edge computing systems (e.g., edge servers, smart Internet of Things (IoT) devices) has exacerbated these traditional trade-offs to a critical point. The concurrent rise of architectural heterogeneity (e.g., big/LITTLE-cores on edge SoCs) and high-contention mixed scenarios (e.g., strict low-latency interactive tasks competing against throughput-bound background computations) means that relying on a single, fixed set of scheduling rules inevitably leads to significant performance compromises. The evolution of the default Linux scheduler from the O(1) scheduler \cite{Love2005LKD}, through the Completely Fair Scheduler (CFS) \cite{Molnar2007CFS}, to the recent EEVDF \cite{EEVDFScheduler}, illustrates that even highly sophisticated algorithms engineered for general-purpose robustness frequently struggle to simultaneously balance real-time responsiveness and background throughput on constrained edge nodes. Consequently, the fundamental scheduling objective must shift from striving to statically \emph{balance} various goals, to \emph{dynamically adapting} which specific goal to prioritize at runtime based on the immediate workload context.

While dynamic adaptability is theoretically achievable within user-space applications \cite{Llumnix2024, DBLatencyHiding2024, Junction2024}, this application-level approach is fundamentally incapable of resolving the scheduling dilemma for complex multitasking edge systems. First, application-level schedulers lack system-wide visibility and are powerless to resolve cross-process physical resource contention (e.g., execution starvation caused by aggressive background threads). Second, forcing individual applications to implement independent dynamic scheduling logic inflates software complexity and introduces redundant runtime execution overhead. Therefore, implementing a unified, dynamic scheduling adaptation mechanism directly for the kernel is not merely an optimization, but a structural necessity for modern edge computing.

Recognizing the necessity of kernel-level adaptation beyond general-purpose baselines (e.g., CFS and EEVDF), existing research has explored alternative formal solutions across different paradigms. However, when applied to dynamic edge computing environments, these specialized approaches encounter significant limitations. First, an extensive line of research focuses on \emph{scenario-specific static policies} \cite{liu1973scheduling, delimitrou2013paragon, kaffes2019shinjuku, mao2025rethinkingprovenancecompletenesslearningbased}, which attempt to optimize performance by pre-selecting fixed algorithmic trade-offs (e.g., universally prioritizing real-time tasks). While effective under predefined configurations, this rigid approach fails to address \textbf{Challenge 1: Achieving dynamic adaptability for complex, unpredictable edge workloads}. Although dynamic adaptation is a universal requirement for modern computing, it is particularly exacerbated at the edge where extreme workload diversity meets strict physical limitations. This rigidity leads to severe performance degradation when the runtime context changes. Second, researchers have investigated \emph{computationally intensive AI-assisted dynamic policies} \cite{downey1997model, delimitrou2014quasar, mao2019learning}, which utilize complex online learning models to calculate optimal schedules continuously. Although highly adaptable, this direction struggles with \textbf{Challenge 2: Confining runtime intelligence within strict embedded resource budgets} \cite{chen2018auto}, as the continuous heavy computation introduces unacceptable runtime overhead and training instability on resource-constrained devices. These limitations imply that a viable scheduling framework must reliably achieve runtime adaptability without relying on prohibitive online computations.

Historically, this scheduling dilemma has been deeply rooted in the kernel's statically compiled scheduling infrastructure, which forces a strict compromise between rigid static policies and high-overhead learning. While recent frameworks mitigate this rigidity via low-overhead eBPF (Linux \texttt{sched\_ext} \cite{Vernet2024SchedExtStatus}) or user-space delegation (e.g., ghOSt \cite{google2021ghost}, Skyloft \cite{jia2024skyloft}), they still lack the \emph{runtime intelligence} for autonomous transitions. Because extreme workload diversity requires fundamentally incompatible queuing disciplines (e.g., the Latency-criticality Aware Virtual Deadline (LAVD) scheduler \cite{Galin1972665}, which dynamically prioritizes interactive tasks based on virtual deadlines to optimize responsiveness, vs. proportional-share trees for throughput) rather than mere parameter tuning, we advocate for \textbf{dynamically switching between a predefined set of lightweight expert policies}, and propose the \emph{Lightweight Adaptive Policy Switcher} (LAPS) to achieve this. To ensure the viability of this dynamic hot-swapping, LAPS completely decouples high-frequency context switching from low-frequency policy transitions. The active lightweight policy dictates \emph{microseconds-level} task arbitration, whereas the switch of policies is strictly limited to macroscopic workload changes. By temporally isolating these operations, LAPS amortizes transitional overheads to maintain negligible task-level adaptation penalty, achieving robust adaptability (Challenge 1) while circumventing severe modeling overheads (Challenge 2).

To actualize this dynamic switching paradigm, LAPS tackles these challenges across two corresponding dimensions. 
First, to guarantee accurate workload state characterization under transient execution noise (\textbf{Challenge 1}), LAPS executes an Observe-Infer-Switch agentic cycle using a lightweight machine learning model and a Time-Weighted Probability Voting Algorithm, augmented by an offline \emph{Phantom Loading} mechanism to rigorously prevent self-induced switching thrashing. 
Second, to eliminate the prohibitive execution overhead of online intelligence computations (\textbf{Challenge 2}), LAPS structurally decouples its decision-making pipeline. Rather than utilizing the ML model to continuously generate or calculate scheduling policies at runtime, LAPS limits its online role strictly to workload classification. All heavy dynamic trade-off calculations are pre-resolved offline to construct a strictly overhead-aware \emph{Policy Mapping Table} that naturally absorbs transient switching penalties. Consequently, the online LAPS runtime executes dynamic adaptation efficiently by simply mapping the inferred workload state to optimal policies via low-cost table lookups.

The main contributions of this research are as follows:

\begin{itemize} \setlength{\itemsep}{0pt}
    \item We design and implement \textbf{Lightweight Adaptive Policy Switcher (LAPS)}, a lightweight, adaptive scheduling framework that intelligently recognizes the runtime workload pattern and dynamically switches to the scenario-optimal scheduling policies.
    \item By constructing the Observe-Infer-Switch cycle, LAPS achieves accurate workload pattern classification using ML-based model, building foundation for scenario-optimal policy matching. It leverages a \textbf{Time-Weighted Probability Voting Algorithm} that uses a voting mechanism with exponential decay to ensure stable and responsive pattern identification, mitigating the impact of transient system noise.
    \item In addition to workload pattern and system statistics, LAPS further takes the run-time policy switching overhead into account. It introduces the \textbf{Phantom Loading Mechanism} to explicitly quantify the switching penalty. By incorporating the simulation, learning, and evaluation of switching overheads into the offline preparation phase, LAPS achieves switching overhead-awareness for its workload pattern classification model and policy mapping table. As a result, LAPS realizes the tradeoff between optimal candidate policy and low runtime switching overhead.
    \item The experimental results show that LAPS significantly outperforms the default scheduler in the vast majority of test scenarios with an improvement of 8.91\%, approaching 98.7\% of the performance of the theoretical optimum. Results also demonstrate that LAPS achieves superior adaptation and robustness across diverse workloads while maintaining strict execution fairness, all with negligible system overhead.
\end{itemize}

\section{Background \& Related Work}\label{sec_background}
This section provides a comprehensive review of the  process scheduling landscape, organized into three key areas. We first discuss the foundational challenges that motivate the need for adaptive scheduling, then survey two dominant alternative paradigms proposed to address the limitations of general-purpose schedulers (e.g., CFS and EEVDF): scenario-specific static policies and AI-assisted dynamic policies. Finally, we examine the recent emergence of extensible scheduling frameworks that provide the technical foundation for our work.

\subsection{Scheduling Challenges in Edge Computing}
Scheduling on modern intelligent edge computing systems faces profound technical and methodological challenges.
First, on the technical front, resource allocation is severely complicated by the rich heterogeneity defining modern embedded architectures. A single scheduling strategy struggles to achieve optimality across all devices, while developing specialized optimization policies for each platform remains an unscalable burden. 
Furthermore, the workload dynamics within these embedded environments have shifted dramatically. Emerging embedded domains, such as smart vehicles, must now handle an increasingly rich variety of task types (e.g., edge AI inference) on top of their inherent real-time requirements. This forces these systems to concurrently execute latency-critical processes and high-throughput background computations.
Assigning the right process to the appropriate computing unit at the optimal time constitutes a significant challenge for general-purpose schedulers like Linux's EEVDF \cite{Galin1972665, qinan2024globalqueue} (which officially succeeded the traditional CFS \cite{Molnar2007CFS} since Linux kernel 6.6 as the default mainline CPU scheduler).

\subsection{Scenario-Specific Static Policies}
To address these challenges, the traditional path in scheduler design involves creating \emph{scenario-specific static policies}. This approach treats scheduling as an engineering problem to be solved for a well-defined context, resulting in highly-optimized schedulers for domains like real-time systems \cite{liu1973scheduling} and virtualized environments \cite{jia2020vsmt}. In the realm of large-scale datacenters, systems such as Shinjuku \cite{kaffes2019shinjuku} and Zygos \cite{prekas2017zygos} have demonstrated that specialized designs can achieve microsecond-scale tail latency. However, these systems generally prioritize high-throughput request-response patterns and often necessitate specific concurrency models or runtime integration, making them less suitable for embedded applications. The design philosophy of specialization persists even as new, flexible kernel frameworks become available, guiding the development of schedulers tailored for specific use cases, such as LAVD for gaming workloads \cite{Galin1972665}. While effective in their target scenarios, the performance of these static policies inevitably degrades when they encounter dynamic, mixed workloads that fall outside their initial design parameters.

This limitation reveals the first primary challenge for edge scheduling:
\textbf{Challenge 1: Achieving Dynamic Adaptability for Complex Edge Workloads.} While the necessity for dynamic adaptability extends across a broad class of computing systems, it becomes an especially prominent bottleneck for edge environments due to the severe collision of highly variable workloads and strict hardware constraints. The scheduling system must dynamically adapt to varying mixed-workloads without sacrificing the low overhead characteristic of static policies.

\subsection{AI-Assisted Dynamic Policies}
The second path driven by advances in machine learning centers on creating a single, \emph{AI-assisted dynamic policy} capable of adapting to any situation \cite{downey1997model, delimitrou2014quasar, lo2015heracles, mao2016resource, mao2019learning, chen2018auto, liu2022sniper, sun2025dynamic}. However, the pursuit of universal adaptability often introduces substantial complexity and overhead. For example, applying reinforcement learning to task distribution can require significant training phases to be effective. Furthermore, addressing resource management in similarly complex domains, such as mixed database workloads, has led to the exploration of highly sophisticated architectures, including those that combine hierarchical deep reinforcement learning with graph neural networks \cite{Xing2025_HDRLDB}. This illustrates a trend where increasing adaptability can correspond with a significant rise in model and system complexity. Most recently, efforts like SchedCP \cite{zheng2025agenticosllmagent} have attempted to leverage Large Language Models (LLMs) to suggest schedulers from shell commands. While innovative, this approach typically functions as a static advisory tool rather than an automated runtime switcher. Moreover, it faces challenges in bridging the semantic gap between textual policy descriptions and their varying physical execution across hardware, alongside the inherent reliability concerns from probabilistic LLM inference.

\begin{figure*}[!htb]
\centering
\includegraphics[width=0.9\textwidth]{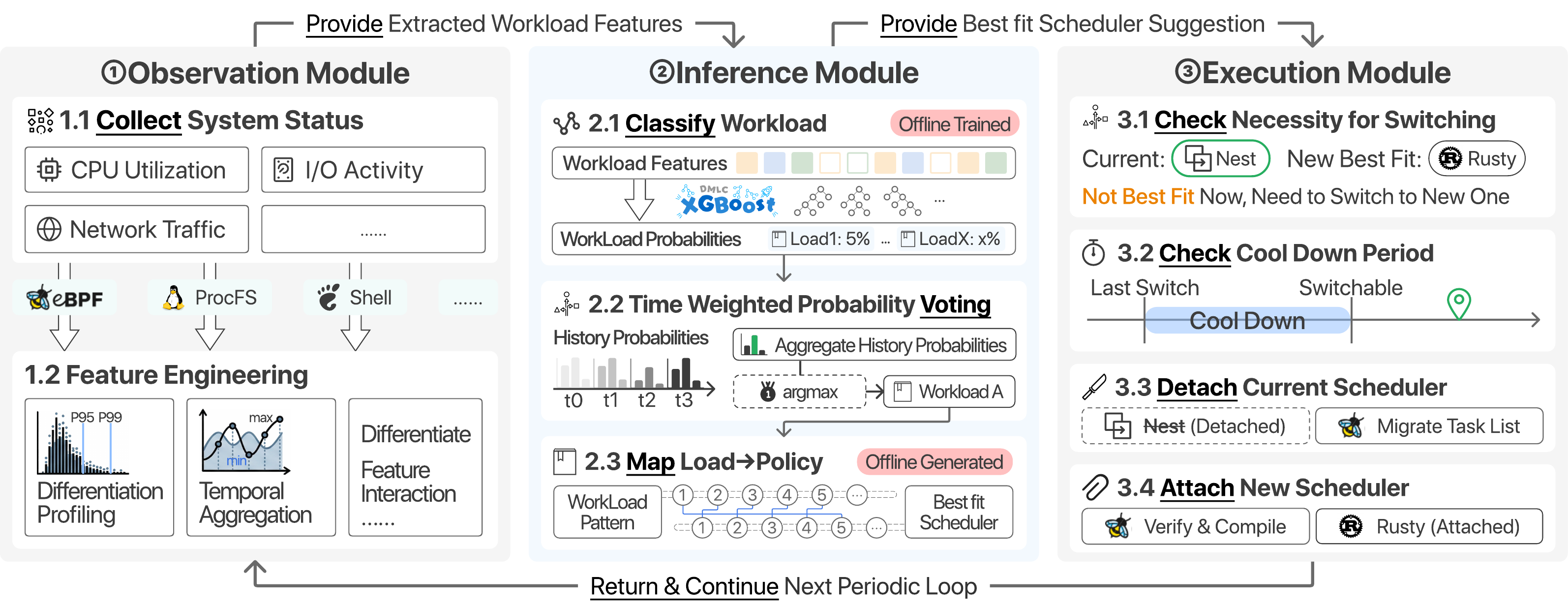}
\caption{Core Framework of the Lightweight Adaptive Policy Switcher. The periodic loop progresses sequentially: (1) \textbf{Observation Module} collects system status to extract workload features; (2) \textbf{Inference Module} classifies the workload and maps it to a best-fit scheduler; and (3) \textbf{Execution Module} evaluates switching necessity and safely attaches the new policy.}
\label{fig:LAPS_framework}
\end{figure*}

This highlights the second critical challenge:
\textbf{Challenge 2: Confining Intelligence within Strict Embedded Resource Budgets.} The scheduling system must provide precise adaptability without exceeding the strict computational and memory constraints typical of edge devices.

\subsection{Rise of Extensible Scheduling Frameworks}
Extensible scheduling frameworks have emerged as a promising direction to overcome these limitations by decoupling scheduling logic from the kernel, thereby enhancing flexibility and accelerating development.
Pioneering work in this area includes Google's \emph{ghOSt} \cite{google2021ghost} and Stanford's \emph{Skyloft} \cite{jia2024skyloft}, which demonstrated the potential of implementing complex scheduling policies outside the core kernel.
However, these early frameworks often require significant, non-standard kernel modifications, which complicates their deployment in production environments.
A key advancement in this domain is the \texttt{sched\_ext} framework, which has been accepted into the mainline Linux kernel \cite{Vernet2024SchedExtStatus}.
As a BPF-based extensible scheduler, \texttt{sched\_ext} provides a standardized, lightweight, and safer mechanism for deploying and dynamically switching scheduling policies, clearing the path for widespread adoption of flexible scheduling solutions \cite{Galin1972665}.

LAPS builds upon the foundation of \texttt{sched\_ext}, but takes a novel approach. While \texttt{sched\_ext} provides the \emph{mechanism} for policy switching, it does not provide the \emph{intelligence} for when and which to switch. Our work fills this critical gap by introducing a lightweight adaptive policy switcher that performs workload-pattern recognition and automated policy selection. By leveraging an expert scheduler set, LAPS achieves system-wide adaptability without the high overhead of complex, monolithic AI-driven schedulers, striking a more effective balance between performance and flexibility.

\section{Framework Overview of LAPS} \label{sec_LAPS}
This section introduces the runtime framework design of LAPS.
Traditional OS schedulers typically operate as mechanical rule executors based on rigid heuristics. While computationally cheap, they are fundamentally incapable of self-adjusting when faced with unpredictable and complex workload changes (\textbf{Challenge 1}). Overcoming this limitation requires a paradigm shift towards an \emph{agent-based architecture}. Recently, intelligent agents have demonstrated remarkable success in complex environments due to their continuous ``Perceive-Reason-Act'' cognitive loops, which enable unparalleled autonomous adaptability to unforeseen state changes. By modeling the scheduling runtime as such a dynamic agent, the system establishes a continuous interaction with its environment rather than relying on static logic. However, realizing this agentic intelligence traditionally demands monolithic, resource-intensive models, strictly conflicting with edge computing constraints. To specifically address \textbf{Challenge 2} (Confining Intelligence within Strict Embedded Resource Budgets), LAPS structurally decouples this agent architecture. By delegating the computationally intensive learning and environment profiling entirely to an offline pipeline (detailed in Section~\ref{sec_prepare}), the online component is distilled into a strictly lightweight agent. This structurally simplified cognitive loop dynamically switches between pre-compiled static policies based on runtime perception of workload changes, achieving agentic adaptability without violating rigorous resource budgets.

As illustrated in Fig.~\ref{fig:LAPS_framework}, the active LAPS runtime executes a streamlined Observe-Infer-Switch cycle, which efficiently instantiates the classic agent architecture through its three modules: the \emph{Observation} module, the \emph{Inference} module, and the \emph{Execution} module.
During runtime, the \emph{Observation} module monitors system workloads via low-overhead eBPF probes. Subsequently, the \emph{Inference} module utilizes a highly compressed, pre-trained classifier to identify workload patterns and retrieves the optimal scheduling policy from a statically generated mapping table. Finally, the \emph{Execution} module performs the switch. This architectural decoupling ensures that intelligent adaptability is achieved without violating the rigorous resource constraints of edge environments.

\subsection{Observation module} \label{sec:observation}
LAPS's intelligence is rooted in the accurate observation of system workloads. The \emph{Observation} module collects system statistics from multiple sources, including kernel-level eBPF programs, the procfs file system, and GNOME Shell for shell layer information. This allows for a comprehensive view of system activity, while maintaining minimal performance overhead. The feature collection process captures multiple dimensions of system behavior:
\begin{itemize}
\setlength{\itemsep}{0pt}
\item \textbf{CPU}: User / System Mode Utilization, Nice / Idle / I/O Wait / Hardware Interrupt / Software Interrupt / Steal Utilization, Hot-Spot Core Ratio
\item \textbf{Memory}: Total / Free / Cached Memory, Total / Available Swap, Buffers
\item \textbf{Disk}: Disk I/O Queue Length, Read/Write Operation Count, Read/Write Latency, Average I/O Size
\item \textbf{Process}: Process Count, GPU-Using Process Count / CPU Utilization, Window-Focused Process CPU / Memory Utilization, Input Event
\item  \textbf{Scheduling}: Task Migration Count, Lock Contention / Acquisition Failure Count, Lock Hold / Thread Blocked Time, Thread Wakeup / Context Switch Latency, Context Switch Count, Run Queue Length
\item \textbf{Network}: Total / New / Closed / Reset Connection Count, Data Sent/Received, Packet Count, Packet Retransmission Rate, TCP/UDP Ratio, Network Interrupt Count / Handling Latency, Data Transmit/Receive Latency, Kernel-Userspace Latency
\end{itemize}

LAPS is not limited to collecting these metrics. Due to the decoupled and modular design of the \emph{Observation} module, integrating new metrics (e.g., device-specific hardware performance counters via PMUs or the \texttt{perf} subsystem) simply requires implementing an additional submodule probe. This allows the system to incorporate new metrics without structural changes to the core inference pipeline.

We adopt a sliding-window-based statistical feature extraction strategy to transform high-dimensional time-series observation data into fixed-dimension feature vectors:

\begin{itemize} \setlength{\itemsep}{0pt}
    \item \textbf{Differentiation and Rate Conversion:} For cumulative counter metrics, we calculate the rate of change per unit time, thus reflecting real-time load intensity.
    
    \item \textbf{Temporal Aggregation:} For state-based metrics, we perform statistical aggregation (Min/Max/Avg) to capture the tendency and range of resource usage levels.
    
    \item \textbf{Distribution Profiling:} For high-frequency latency data, we extract key quantiles ($P_{50}$, $P_{95}$, $P_{99}$), mean, and standard deviation. This preserves the non-Gaussian distribution characteristics and captures tail anomalies.
    
    \item \textbf{Feature Interaction and Construction:} We construct high-order features based on domain knowledge. For instance, we use the Gini coefficient to quantify imbalances (e.g., in multi-core loading) and dimensionless ratios to decouple absolute numerical magnitudes. Through these extensive feature interactions and constructions, LAPS further extracts complex computational properties, explicitly capturing the fundamental nature and arrival patterns of the workloads.
\end{itemize}

Through this feature extraction process, we obtain a 150-dimensional feature vector, which serves as the input for both offline model training and runtime inference in the subsequent inference module.

\subsection{Inference module}

The Inference module is the intelligent core of LAPS, responsible for analyzing collected features and identifying workload patterns. This module employs a machine learning-based classification model that has been trained offline on diverse workload scenarios to recognize common patterns such as CPU-intensive computing, I/O-bound operations, interactive applications, and mixed workloads. The classifier algorithm is based on XGBoost, chosen for its excellent performance on structured data and ability to handle feature interactions. Through these extensive feature interactions and constructions, LAPS further extracts complex computational properties, explicitly capturing the fundamental nature and arrival patterns of the workloads.

To address the challenge of temporal stability in dynamic environments, LAPS implements a Time-Weighted Probability Voting mechanism to aggregate classification results over a sliding window. It ensures that scheduling decisions are based on stable workload patterns rather than transient fluctuations. The design of this voting mechanism was carefully evaluated against alternatives to strike an optimal balance between stability and agility. A standard \textit{majority voting} scheme, for example, retains only the Top-1 predicted label at each tick, aggressively discarding probability confidence information. Under transient execution noise, this can erroneously magnify tiny probability gaps into definitive wrong labels, potentially triggering unwarranted and expensive policy switches. Conversely, \textit{hysteresis-based switching} relies on rigid deadbands that prevent oscillation but react sluggishly to genuine, sudden workload transitions. In contrast, our Time-Weighted Probability Voting employs soft voting combined with an exponential decay. This design retains subtle confidence differences over the window, giving the highest weight to the current cycle while leveraging historical distributions to smooth out transient noise and break ties effectively. The detailed procedure is formalized in Algorithm~\ref{alg:voting}.
This approach effectively filters out transient system fluctuations while remaining responsive to genuine workload transitions.

\begin{algorithm}
\caption{Time-Weighted Probability Voting}
\label{alg:voting}
\begin{algorithmic}[1]
    \Statex \textbf{Input:} Sequence of probability distributions \{$P_1, \dots, P_W$\} over a window $W$, threshold $\theta$, decay factor $\alpha$.
    \Statex \textbf{Output:} Final selected class $C_{final}$.
    \State Initialize aggregated scores $A_c = 0$ for all classes $c$.
    \For{$t = 1$ to $W$}
        \For{each class $c$ with probability $p_{t,c}$ in $P_t$}
            \State Let $p'_{t,c} = \begin{cases} p_{t,c} & \text{if } p_{t,c} > \theta \\ 0 & \text{otherwise} \end{cases}$
            \State $A_c \leftarrow A_c + p'_{t,c} \cdot \alpha^{W-t}$
        \EndFor
    \EndFor
    \State $C_{final} \leftarrow \operatorname{arg\,max}_c A_c$
    \State \textbf{return} $C_{final}$
\end{algorithmic}
\end{algorithm}

Once the stable workload classification result is determined, the Inference module queries its internal mapping table to select the scenario-optimal scheduling policy. Note that the workload classifier model and this policy mapping table are both generated through a specialized offline preparation pipeline, which will be discussed in detail in Section~\ref{sec_prepare}.

\subsection{Execution module}

The Execution module operates on a \emph{seconds-level} periodic basis to dynamically execute the optimization of system process scheduling. During each check, the Execution module first checks the currently active scheduling policy and retrieves the timestamp of the last switch. To prevent system oscillation, a policy switch is triggered only when the Inference module suggests a new optimal strategy and the elapsed time since the last operation exceeds a preset cooldown period.

During the switch process, the Execution module first unloads the currently active custom scheduler. This action triggers the kernel's task draining mechanism, which seamlessly migrates all tasks managed by the outgoing policy back to the Linux kernel's default \texttt{EEVDF} scheduler. This fallback mechanism ensures that system services and process execution remain uninterrupted during the switch gap between the cessation of the old policy and the activation of the new one.

Subsequently, LAPS initiates the loading procedure for the target policy. At this stage, the kernel's BPF verifier intervenes to perform code safety checks and Just-In-Time(JIT) compilation on the new scheduling logic. Only after passing verification is the new BPF program attached, formally taking over the task queue and scheduling authority.

Given the inherent overheads associated with task migration and BPF verification, unrestricted frequent switching would degrade system performance, even if the selected policies were theoretically optimal. Therefore, the Execution module enforces a mandatory cooldown period to strike a balance between policy dynamism and system stability. In practice, due to inherent application loading and human operation times, users typically do not shift workloads at ultra-high frequencies. Therefore, this cooldown mechanism primarily serves as a necessary safeguard for extreme corner cases, rather than a frequent restriction during typical system operation.

\section{Offline Preparation} \label{sec_prepare}
\begin{figure*}[htb]
\centering
\includegraphics[width=0.9\textwidth]{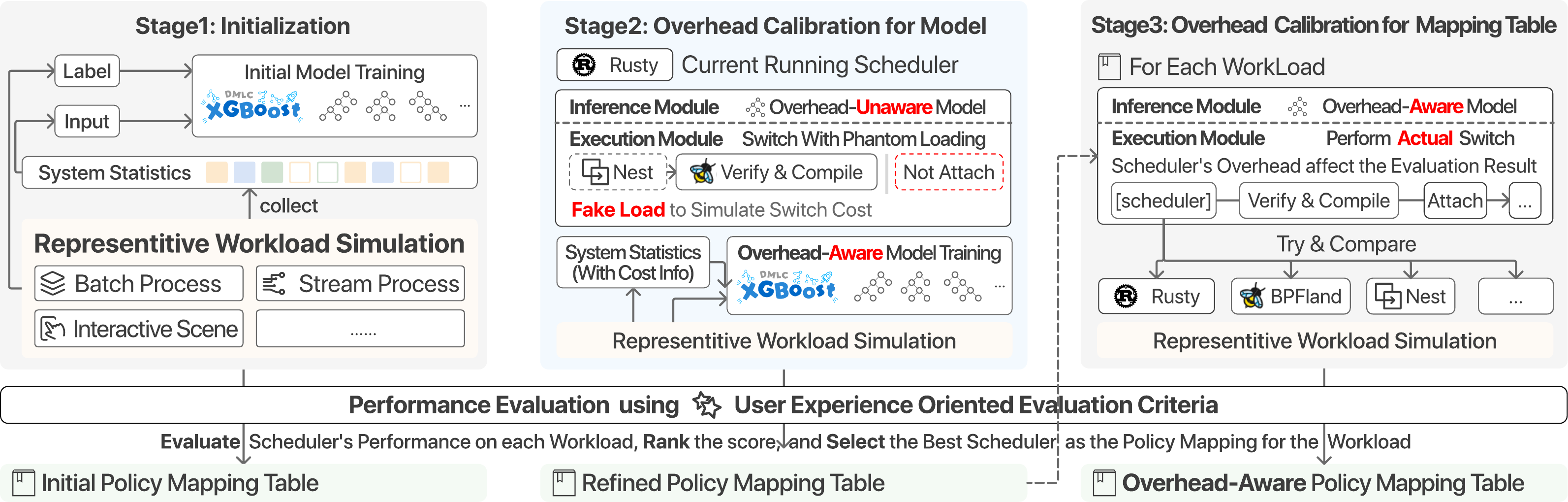}
\caption{The Workflow of Offline Preparation stages.}
\label{fig:offline_cycle}
\end{figure*}

This section introduces the offline preparation of LAPS's core intelligence: the Workload Classifier Model and the Policy Mapping Table. 
To rigorously address \textbf{Challenge 1} (Achieving Dynamic Adaptability for Complex Edge Workloads), LAPS must establish a precise mapping between transient workload states and optimal static scheduling policies. Constructing this mapping presents two technical hurdles: accurate labeling requires a unified scoring strategy to compare multi-metric performance on edge devices, and the decision logic must account for the non-negligible operational overheads inherent in policy switching (e.g., eBPF JIT compilation).

To resolve these hurdles, we design a systematic offline preparation pipeline. We first introduce a metric normalization method to seamlessly aggregate multi-scenario, multi-metric performance into a comparable index. Building on this foundation, we propose a Three-Stage Offline Preparation Strategy (Fig.~\ref{fig:offline_cycle}). This process evolves from Initialization to Overhead Calibration for the model and Mapping Table, ensuring that the finalized classifier model and mapping table provide robust profitability in realistic edge environments containing switching overheads. Finally, we demonstrate that LAPS facilitates rapid adaptation to diverse new hardware by simply executing the Stage 3 calibration on target platforms.

\subsection{Metric Normalization}
\label{sec:eval_criteria}
To comprehensively profile a scheduler's versatility, it must be evaluated across a wide spectrum of heterogeneous scenarios (e.g., interactive gaming, background compiling). Each scenario is characterized by multiple distinct metrics (e.g., frame rate, I/O throughput, tail latency). A core challenge in this multi-scenario, multi-metric evaluation is that direct aggregation of raw metrics is mathematically invalid due to their disparate units and magnitude scales (e.g., summing milliseconds and MB/s). Furthermore, simple system-wide summaries (e.g., CPU utilization) fail to capture the detailed performance bottlenecks of specific interactive workloads. To resolve this, we map all raw metrics into a standardized, dimensionless $[0, 1]$ score interval.
Instead of linear Min-Max normalization, which is highly vulnerable to worst-case outliers (especially for long-tail latencies), we adopt a robust sigmoid-based transformation. This softly bounds extreme tail values and ensures that no single unbounded metric disproportionately dominates the composite score. The normalized score combines a stability score and an absolute value score:
\begin{equation}
    S(x) = \frac{1}{1 + \sigma^2} + \frac{1}{1 + e^{\pm (\frac{x-\theta}{f})^{2}}}
\end{equation}

Here, $\sigma$ denotes standard deviation. The threshold $\theta$ is set per metric, and the scale factor $f$ is chosen from the metric's empirical dispersion. To consolidate the individual metric scores ($S(x)$) into a final evaluation without introducing subjective bias, we employ a hierarchical, equal-weight aggregation strategy. First, within a given workload scenario, metrics are categorized into functional groups (e.g., latency metrics, availability metrics, throughput metrics). The normalized scores are averaged within each group to prevent scenarios with a larger sheer number of a certain metric type from dominating the outcome. Then, group averages are equally combined to produce a unified scenario score. Finally, the overall scheduler performance is calculated as the arithmetic mean across all evaluated scenarios. This parameter-free aggregation ensures impartial profiling by treating every metric group and scenario equitably.

\subsection{Classifier Model \& Mapping Table Generation}
This part introduces the 3-stage strategy realizing the process of model updating from a coarse-grained level to a refined level. 

\subsubsection{Stage 1: Initialization}

This stage primarily focuses on training LAPS's initial workload recognition model with data collected in a standardized test environment and establishing a preliminary mapping between workload patterns and optimal policies.
To accomplish this, we first deploy different Linux schedulers to run completely through a series of pre-defined, typical test scenarios to simulate particular workloads. During this process, two key types of data are collected simultaneously: 1) System Operational Statistics Data, as detailed in Sec~\ref{sec:observation}, which forms the feature vectors for training the workload classification model, and 2) Simulated Scenario Performance Metrics that will later be assessed with the metric normalization method (Sec~\ref{sec:eval_criteria}) to objectively score and rank each policy.
With the system operational statistics, we train an initial workload classification model (detailed training hyperparameters for reproducibility are provided in Appendix A. Concurrently, based on the performance evaluation results, we determine the optimal scheduling policy for each workload, thereby constructing an initial policy mapping table of ``scenario-optimal policy''.

In our experiments, the expert set in LAPS's policy mapping table includes: 
\emph{scx\_p2dq}\cite{scxp2dq}, {\emph{scx\_bpfland}\cite{scxbpfland}}, {\emph{scx\_nest}\cite{lawall2022scheduling}}, {\emph{scx\_lavd}\cite{scxlavd}}, {\emph{scx\_simple}}, {\emph{scx\_flash}\cite{scxflash}}, {\emph{scx\_rusty}\cite{scxrusty}}, and the default {\emph{EEVDF}~\cite{EEVDFScheduler}}. 
We curated this diverse set to demonstrate LAPS's capability to synthesize the strengths of distinct scheduling paradigms. 
However, \textbf{LAPS is by no means restricted to these options}. Any scheduler implemented via \texttt{sched\_ext}, or essentially any policy capable of dynamic hot-loading and unloading via command-line interfaces, can be seamlessly integrated as an expert.

\subsubsection{Stage 2: Overhead Calibration for Model}

A critical challenge in dynamic scheduling is that the act of switching policies itself introduces non-negligible overhead—primarily from eBPF verification and JIT compilation—which can perturb system metrics and lead to decision instability. Depending on the specific implementation, the policies' overhead may also appear different. To address this, we introduce a \textit{Phantom Loading Mechanism} to explicitly quantify the transient switching penalty without introducing the confounding behavioral shifts of a new scheduling logic.

\begin{listing}[t]
\begin{lstlisting}[language=C]
// Phase 1: Incur load & verification overhead
g_skel = SCX_OPS_OPEN(simple_ops, scx_example);
SCX_OPS_LOAD(g_skel, simple_ops, scx_example, uei);
if (is_phantom_mode){
    while (!exit_req) { sleep(1); }
} // Unreachable (below): scheduler never activates
SCX_OPS_ATTACH(g_skel, simple_ops, scx_example);
\end{lstlisting}
\caption{Phantom Mode Loading Example.}
\label{lst:phantom}
\end{listing}

As shown in Listing~\ref{lst:phantom}, we modify the scheduler lifecycle to support a Phantom Mode. In this mode, the scheduler executes the complete loading and verification sequence (invoking \texttt{SCX\_OPS\_LOAD}) but \textbf{halts} before the final attachment phase (\texttt{SCX\_OPS\_ATTACH}). 
This allows LAPS to inject the realistic latency and CPU overhead of a policy switch into the system without actually altering the active scheduling policy.
During this stage, LAPS operates using the initial mapping from Stage 1. However, when triggering a switching decision, it invokes the target policy in Phantom Mode to collect overhead-inclusive training data. This calibration is imperative to prevent self-induced thrashing. In dynamic scheduling, executing a policy switch inherently pollutes the observable system state. If left uncalibrated, the observation module would erroneously interpret the transient latency and CPU spikes of JIT compilation as a dense computational workload, triggering immediate, reciprocal scheduler switches. Phantom Loading solves this by training the model to recognize and disregard self-inflicted system perturbations, endowing it with the rigorous cognitive stability needed to differentiate actual workload dynamics from mere scheduling artifacts.

\subsubsection{Stage 3: Overhead Calibration for Mapping Table}

In this final stage, LAPS operates in a fully functional mode with actual policy switching enabled. Using the Stage 2 mapping as a baseline, we conduct a systematic profiling sweep. For each workload category, we iteratively override the default scheduling mapping config, forcing the system to evaluate every candidate policy on such workload while keeping the rest of the configuration constant.
This process generates a comprehensive dataset linking workload patterns to runtime performance. Since LAPS is performing switches, the recorded metrics capture the net benefit of each policy, weighing the policy's optimization benefits against its inherent operational and switching overheads. By training on this local data, the final model and mapping table are calibrated against the specific hardware and software environment, ensuring that decisions are fully overhead-aware now. 

Crucially, this stage is designed to be incremental and opportunistic. It can operate during idle periods. Furthermore, it also allows for expert scheduler set extensibility: introducing a new policy requires a differential profiling run, without the need to re-evaluate the entire scheduler set.

\subsection{Fast Adaptation to New Hardware}

Deploying the LAPS onto a new hardware platform is a streamlined process. By exclusively running the Stage 3 on the target machine, we can directly gather statistics of the new environment and its available policies, thus generating a precise, hardware-specific policy mapping table and finetuning the classification model to adapt to the new platform.

\section{Evaluation} \label{sec_evaluation}
This section systematically evaluates the efficacy of LAPS. 
We first detail the experimental settings in Section~\ref{sec:eval_setup}. %
In Section~\ref{sec:eval_claim1}, we validate LAPS's ability to handle complex mixed-workloads and resolve real-time interactive bottlenecks on constrained edge nodes. 
In Section~\ref{sec:eval_claim3}, we verify LAPS's adaptability by evaluating its performance on diverse and heterogeneous environments. 
In Section~\ref{sec:eval_claim4}, we confirm that LAPS maintains system robustness even in the presence of mispredictions, and presents fairness to background tasks, indicating that LAPS provides a safe optimization path for resource-constrained devices.
In Section~\ref{sec:eval_ablation}, we conduct ablation studies to validate the efficacy of LAPS's key design components. 
Finally, Section~\ref{sec:eval_overhead} analyzes LAPS's system overhead to confirm its practicality, and Section~\ref{sec:eval_summary} summarizes the experimental results.

\begin{table}[t]
\centering
\caption{Evaluation Platform Configuration}
\label{tab:vm_configurations}
\resizebox{\linewidth}{!}{
\begin{tabular}{c|c|c|c|r|c}
\toprule
\textbf{ID} & \textbf{BaseCPU} & \textbf{Cores} & \textbf{Thrd.} & \textbf{Mem} & \textbf{Proto.} \\
\midrule
\multicolumn{6}{c}{\textbf{Virtual Machine Environments}} \\
\midrule
VM1 & i5-6500  & 4 & 4 & 8G & \checkmark \\
VM2 & i5-6500  & 4 & 4 & 16G & \checkmark \\
VM3 & i5-9400  & 6 & 6 & 12G & \ding{55} \\
VM4 & i5-9400  & 6 & 6 & 16G & \ding{55} \\
VM5 & i5-9400  & 4 & 4 & 8G & \ding{55} \\
VM6 & i5-9400  & 2 & 2 & 8G & \checkmark \\
VM7 & i7-12700 & 8P+4E & 20 & 16G & \ding{55} \\
VM8 & i7-12700 & 8P & 16 & 16G & \ding{55} \\
VM9 & i7-12700 & 6P & 12 & 16G & \checkmark \\
VM10& i7-12700 & 4P+4E & 12 & 16G & \ding{55} \\
\midrule
\multicolumn{6}{c}{\textbf{Physical Machine (Bare Metal)}} \\
\midrule
BM1 & i7-12700 & 8P+4E & 20 & 16G & \ding{55} \\
\bottomrule
\end{tabular}
}
\end{table}

\begin{table}[t]
\centering
\caption{Benchmark Scenarios Matrix}
\label{tab:scenarios_matrix_academic}
\small
\renewcommand{\arraystretch}{0.5}
\begin{tabular}{c|c|c|c|c}
\toprule
 & \textbf{\makecell[c]{Game \\ Play}} & \textbf{\makecell[c]{Office File \\ Editing}} & \textbf{\makecell[c]{Web \\ Browsing}} & \textbf{\makecell[c]{Audio \\ Remix}} \\
\hline
\textbf{\makecell[c]{Archive \\ Extraction}} & S1 & S8  & S15 & S22 \\
\hline
\textbf{\makecell[c]{Blender \\ Render}}     & S2 & S9  & S16 & S23 \\
\hline
\textbf{\makecell[c]{Kernel \\ Compile}}     & S3 & S10 & S17 & S24 \\
\hline
\textbf{\makecell[c]{LLM \\ Generate}}       & S4 & S11 & S18 & S25 \\
\hline
\textbf{\makecell[c]{Disk \\ I/O}}    & S5 & S12 & S19 & S26 \\
\hline
\textbf{\makecell[c]{Network \\ Transfer}}   & S6 & S13 & S20 & S27 \\
\hline
\textbf{\makecell[c]{Video \\ Render}}       & S7 & S14 & S21 & S28 \\
\bottomrule
\end{tabular}
\end{table}

\subsection{Experimental Setup}\label{sec:eval_setup}

\subsubsection{Evaluation Platform Configurations}

Consistent with LAPS's design as a lightweight local scheduler operating within each OS, we deployed LAPS to 11 standalone environments separately to assess its adaptability under varying resource constraints.

Our automated evaluation spans 10 Virtual Machines (VMs) and one physical bare-metal (BM) node, rigorously simulating the hardware diversity of modern automotive and edge computing environments. The evaluation comprises three key dimensions:
\begin{itemize}
    \item \textbf{Embedded Topology Simulation:} Our platform selection reflects the current industrial shift toward high-performance centralized compute in edge and automotive domains, where high-performance compute platforms are increasingly adopted (e.g., Tesla's deployment of Intel Infotainment processor\cite{tesla_nhtsa_2024_intel} and AMD Ryzen\cite{amd_computex_2021_tesla} in MCUs). To evaluate LAPS across this evolving hardware landscape, our environments span a wide spectrum of computational capacities from 2 to 20 threads (Table~\ref{tab:vm_configurations}). This configuration emulates both localized IoT nodes and high-end automotive MCUs, while the inclusion of P/E hybrid cores in VM7 and VM10 further verifies LAPS's adaptability to big/LITTLE heterogeneous topologies typical of embedded SoCs.
    \item \textbf{Interference Isolation:} We enforce hardware-level isolation by dedicating an entire physical node to a single active VM during tests. Coupled with strict CPU core pinning (vCPU to physical core mapping), this setup guarantees a pure baseline devoid of scheduling noise. We adopted this strict environment management protocol to explicitly address the profound impact of OS-level noise on scheduling evaluations. By strictly isolating the VM, disabling automatic background system updates, and dedicating physical cores, we ensure that the comparative performance among different schedulers is not biased by uncontrolled resource interference, context switching artifacts, or hypervisor-induced jitter across different experimental runs. This guarantees consistent, reproducible, and highly objective evaluation conditions.
    \item \textbf{Cross-Device Adaptability:} To validate practical deployment robustness, we calibrate LAPS on a subset of \textbf{prototype devices} (VM 1, 2, 6, 9) and evaluate its zero-shot generalization on the remaining seven \textbf{adapted devices}, explicitly testing its transition across new processor architectures and from virtual to bare metal.
\end{itemize}

\subsubsection{Benchmark Settings}

\begin{figure*}[t]
\centering
  \includegraphics[width=\textwidth]{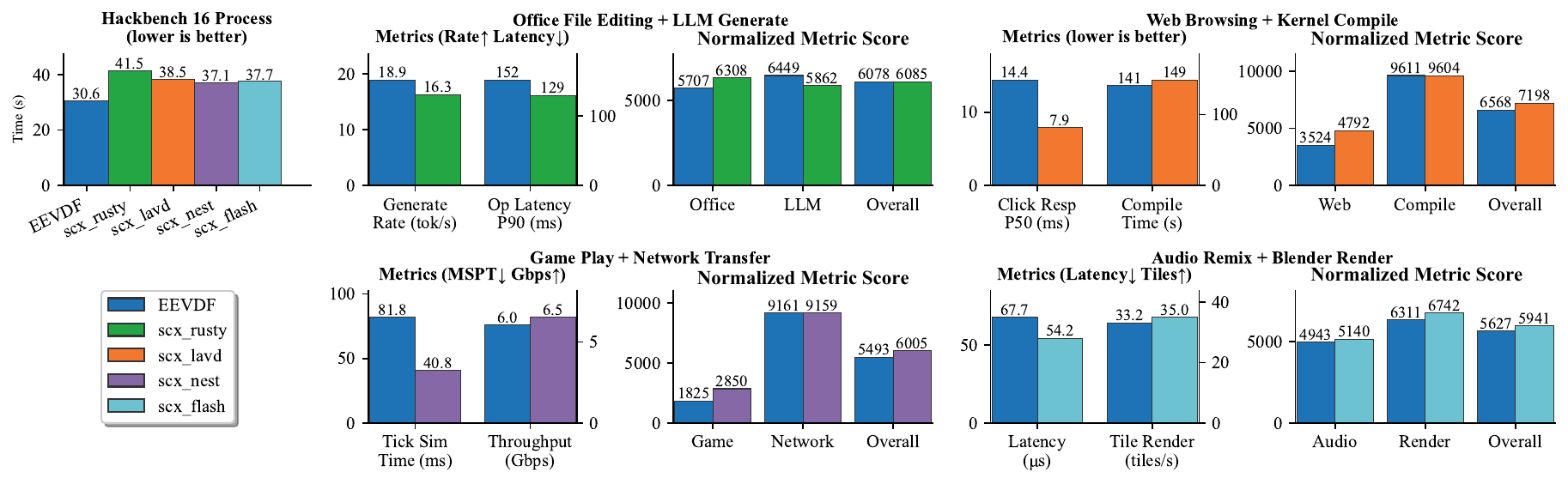}
  \caption{Discrepancy between throughput-oriented synthetic benchmark (Hackbench\cite{Ansen88_hackbench_v2.3}) and real-world application interactivity metrics. Our normalized composite score enables objective comparability across diverse multi-task scenarios. Scores are scaled by $10^4$ for readability.}
  \label{fig:hackbench_vs_metrics}
\end{figure*}

\quad To simulate the resource contention patterns of real-world applications in smart vehicle cabins and other edge devices, we constructed a benchmark suite of 28 scenarios (detailed design and configurations are provided in Appendix B. 
As shown in Table~\ref{tab:scenarios_matrix_academic}, these scenarios are generated by pairing 4 interactive, latency-sensitive applications with 7 resource-intensive, background workloads. Our benchmark suite is meticulously built on widely-adopted open-source applications that are highly representative of diverse in-vehicle and edge-computing scenarios. To ensure reproducible and synchronized execution across all tests, we developed a sophisticated upper-level automation and orchestration framework. Specifically, the framework automatically prepares each scenario to ensure a stable and comparable baseline state, and strictly synchronizes the concurrent start timing to evaluate performance under realistic resource contention. Upon test completion, the framework systematically collects the application-level performance metrics from all involved scenarios. The applications are cross-combined by category to reduce benchmark bias and dramatically improve the representativeness of the evaluation.

\begin{figure*}[t]
\centering
  \includegraphics[width=\textwidth]{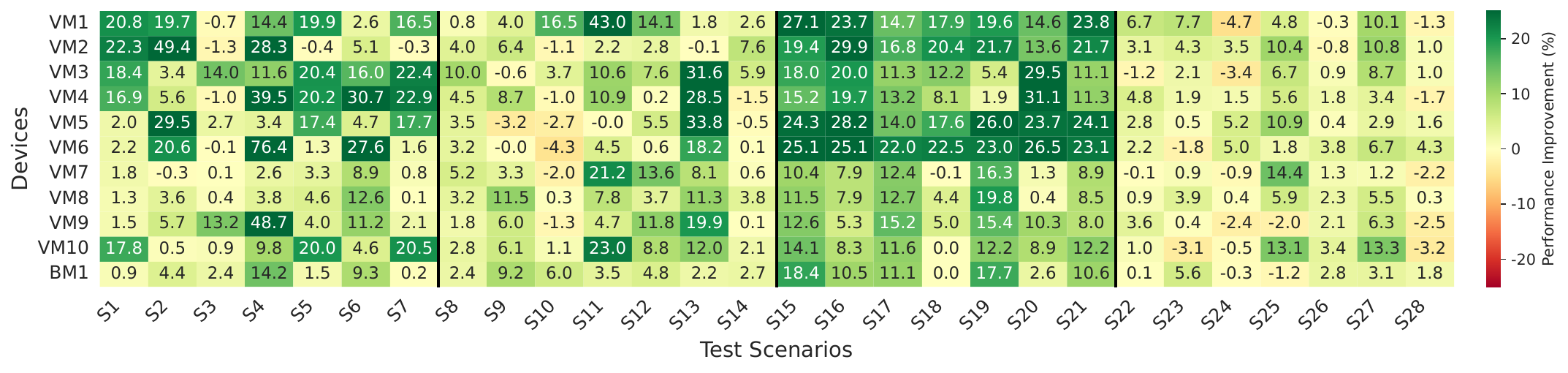}
  \caption{Performance comparison between LAPS and EEVDF (scenarios × devices)}
  \label{fig:main_heatmap}
\end{figure*}

\subsubsection{Baselines and Metrics}

To comprehensively evaluate LAPS, we deployed eight baselines ranging from the mainline kernel standard to other specialized \texttt{sched\_ext} policies. 
We selected these schedulers to cover the spectrum of conflicting optimization goals that modern systems must balance, such as latency, throughput, fairness, and hardware locality. To ensure absolute fairness in our comparative evaluation, all expert \texttt{sched\_ext} schedulers were strictly configured with their default or officially recommended parameters to reflect their intended general-purpose usage and avoid introducing any artificial tuning bias favoring our proposed approach. Furthermore, it is worth noting that in embedded systems, traditional real-time scheduling classes (e.g., \texttt{SCHED\_FIFO}, \texttt{SCHED\_RR}, and \texttt{SCHED\_DEADLINE})and classical adaptive frameworks (e.g., AQuoSA\cite{palopoli2009aquosa}, which shares the same Earliest Deadline First and Constant Bandwidth Server (EDF/CBS) foundation) are typically reserved for a small, static set of critical threads requiring manual, task-specific tuning. While these classes are vital for such real-time threads, they are not designed for globally managing complex, dynamically mixed workloads. Attempting to statically map our diverse multi-task scenarios to real-time parameters would introduce arbitrary tuning assumptions and fail to reflect typical OS scheduling practices. Consequently, these real-time policies fall outside the scope of this general-purpose evaluation, and the chosen expert schedulers will serve as our primary baselines.

\begin{itemize} \setlength{\itemsep}{0pt}
\item \textbf{Mainline Standard:} EEVDF\cite{EEVDFScheduler} serves as the primary point of comparison. As the default mainline Linux scheduler since kernel 6.6, EEVDF represents the latest state-of-the-art in monolithic, ``one-policy-fits-all'' static optimization. Because EEVDF was specifically designed to provide tighter latency guarantees by introducing virtual deadlines, utilizing it as our baseline provides a more rigorous and modern calibration reference.
\item \textbf{Latency-Critical Policies:} To represent policies optimized for responsiveness and tail latency, we deployed \textbf{scx\_lavd}\cite{scxlavd} (based on virtual deadline) and \textbf{scx\_bpfland}\cite{scxbpfland} (based on identified interactive priority) to challenge LAPS in interaction-intensive scenarios. %
\item \textbf{Topology and Hardware-Aware Policies:} To represent throughput-oriented and topology-aware optimizations, we deployed \textbf{scx\_p2dq}\cite{scxp2dq} and \textbf{scx\_nest}\cite{lawall2022scheduling}. P2DQ uses the classic Power of Two Choices algorithm\cite{mitzenmacher2002power} to achieve load balancing across Last Level Caches (LLCs) and Non-Uniform Memory Access (NUMA) nodes. Nest targets frequency-aware core selection. Comparison with these policies can test LAPS's ability to leverage hardware-specific traits.
\item \textbf{Fairness and Predictability:} We deployed \textbf{scx\_flash}\cite{scxflash} (Earliest Deadline First-based) and \textbf{scx\_simple}\cite{eunomia_scx_simple} (virtual time-based) to represent classic algorithms that prioritize strict fairness and algorithmic predictability over heuristic tuning.
\item \textbf{Architectural Hybrid:} The \textbf{scx\_rusty}\cite{scxrusty} represents user\&kernel-space hybrid models, evaluating LAPS against policies that offload decision-making logic to user-space.
\end{itemize}
This diverse baseline setting enables us to evaluate LAPS along two critical dimensions. 
First, domain-specific optimality. We examine if LAPS can effectively identify workload patterns and approximate the performance of the best-in-class scheduler for that specific domain (e.g., scx\_lavd for gaming).
Second, adaptation ability. We assess LAPS's robustness in complex environments. Unlike baselines that are optimized for a narrow scope, we examine if LAPS can consistently deliver high performance across diverse workloads, effectively acting as a superset of the underlying expert policies.

To comparably evaluate the schedulers, we utilize the normalized composite score defined in Section~\ref{sec:eval_criteria}. Direct performance comparison across mixed concurrent workloads is inherently challenging without rigorous mathematical normalization. This is because the performance metrics of different workloads vary significantly in their units, scales, and optimization directions (e.g., Game Play aims to maximize frame rate and minimize frame time, whereas LLM Generate aims to minimize the time to first token and maximize token generation rate). Alternative approaches, such as direct absolute-value summation, are often compromised because the final score can be dominated by numerically large metrics (e.g., network bytes transferred), thereby losing physical meaning and skewing the evaluation. Similarly, purely rank-based scoring methods discard the magnitude of the performance differences, treating a small incremental improvement identically to a major speedup. Our normalized score robustly standardizes these disparate metrics into a unified, dimensionless scale. This mathematically sound approach preserves the relative performance gains, ensures objective comparability across heterogeneous applications, and fairly captures the true end-to-end interactivity improvements perceived by the user.

As Figure~\ref{fig:hackbench_vs_metrics} highlights, classical synthetic benchmarks like Hackbench often diverge from practical application responsiveness. EEVDF's superior Hackbench outcome fails to reflect proportional improvements in end-to-end interactive scenarios. This inconsistency highlights the limitations of synthetic benchmarks for complex edge workloads. Therefore, our evaluation strictly focuses on aggregating multi-dimensional metrics (e.g., latency, jitter, frame rate) across diverse realistic scenarios. 

\begin{figure*}[t]
\centering
  \includegraphics[width=0.97\textwidth]{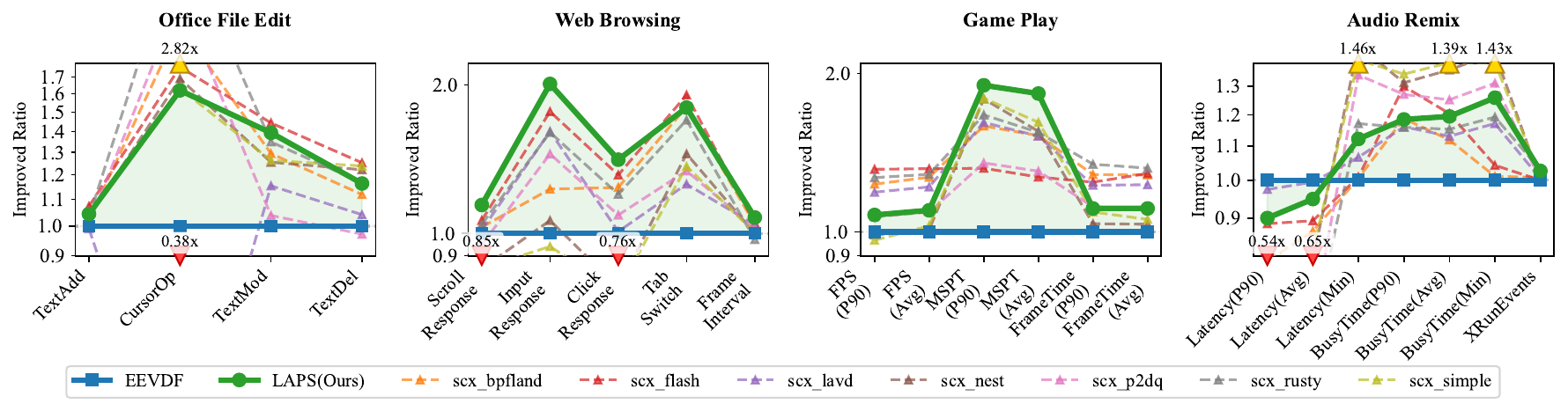}
  \caption{Parallel coordinates plot showing average improvement ratios across different devices. LAPS demonstrates the ability to approach Pareto optimality by exploiting the diverse strengths of existing schedulers based on the underlying hardware. }
  \label{fig:parallel_coords}
\end{figure*}

\begin{figure*}[t]
\centering
  \includegraphics[width=0.97\textwidth]{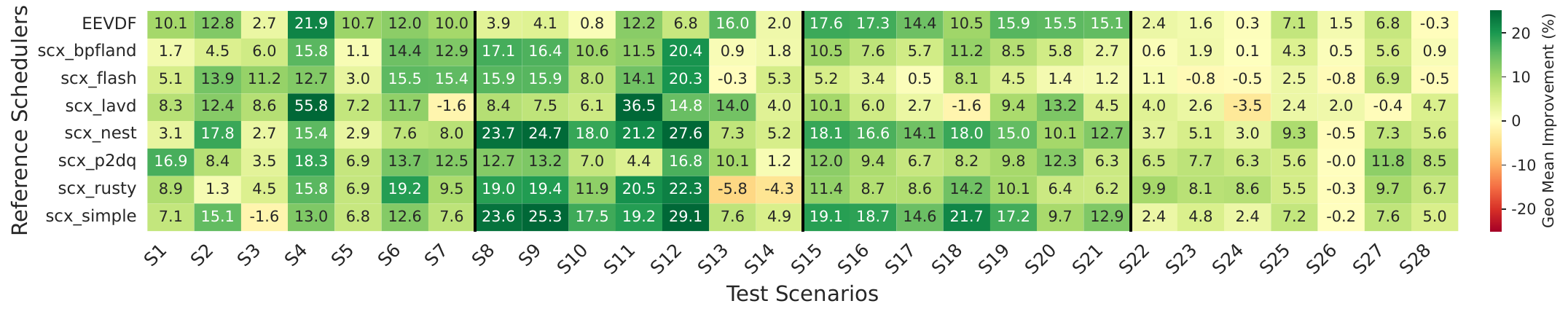}
  \caption{Performance comparison between LAPS and other reference schedulers. The heatmap illustrates that LAPS outperforms or matches static policies across diverse scenarios and devices. }
  \label{fig:heatmap_geomean}
\end{figure*}

\subsection{Case Study I: Handling Complex Mixed-Workloads on Edge Nodes}\label{sec:eval_claim1}
To demonstrate how LAPS resolves \textbf{Challenge 1} (Achieving Dynamic Adaptability for Complex Edge Workloads), we evaluate its performance under high-contention mixed scenarios. 

\textbf{Superiority Over Default Baseline.} A core premise of LAPS is that the dynamic policy selection paradigm mitigates the performance limitations of static policies across diverse workloads. Figure~\ref{fig:main_heatmap} substantiates this by comparing LAPS's performance against EEVDF. The results show that LAPS outperforms EEVDF in 87.0\% of all scenarios, with a global average improvement of +8.91\% with 95\% confidence interval (CI) ranging from 7.75\% to 10.07\%. 
This highlights a fundamental vulnerability of single schedulers on edge nodes: a policy optimized for one type of workload cannot guarantee optimal performance for another. 
In contrast, LAPS overcomes this by switching to the most suitable scheduler on demand.

Notably, in some scenarios LAPS does not significantly outperform EEVDF. This is because EEVDF itself is the optimal or near-optimal choice for that specific workload. In these cases, LAPS correctly identifies this and selects EEVDF (or a similarly performing scheduler), with its marginal overhead sometimes resulting in performance slightly below the native EEVDF. This is not a failure of LAPS's logic but a reflection of the current scheduler set's boundaries. As the portfolio of expert schedulers expands, the performance gains of LAPS are expected to increase further.

\textbf{Metric Trade-offs Mitigation.}
The necessity of LAPS's dynamic approach is further demonstrated in Figure~\ref{fig:parallel_coords}. 
The plot exposes the tradeoff inherent to static policies: aggressive optimization in one area usually precipitates a collapse in another. 
A prime example is the \emph{Audio Remix} workload, where some baseline schedulers boost individual metrics to $\sim1.4\times$ but suffer a catastrophic drop to $0.54\times$ in tail latency metrics ($P_{90}$). 
Unlike static policies that maximize specific metrics, LAPS adopts a holistic optimization strategy. It manages to achieve metric equilibrium and foreground-background fairness, effectively rejecting policies that offer marginal gains in one metric at the cost of starvation in another. It ensures that foreground responsiveness doesn‘t come at the expense of system stability or background progress, validating it as the optimal choice under complex trade-offs.

\textbf{Superiority Over Specialized Static Schedulers.}
Figure~\ref{fig:heatmap_geomean} further validates this by breaking down LAPS's relative performance against each specific baseline. The heatmap presents significant improvement of LAPS under various scenarios, confirming that LAPS effectively leverages the diverse capabilities within its scheduler set. 
Specifically, when compared to Latency-Critical Policies (e.g., \emph{scx\_lavd}), LAPS achieves substantial gains in high-saturation scenarios (e.g., S4, S12). This indicates that while baseline policies struggle to maintain responsiveness under heavy background contention (such as LLM generation or intensive Disk I/O), LAPS effectively isolates the foreground tasks to preserve interactivity. 
Similarly, against Topology and Throughput-Oriented Policies (e.g., \emph{scx\_nest} and \emph{scx\_p2dq}), LAPS demonstrates comprehensive superiority across both Gaming (S1-S7) and Office (S8-S14) workloads. While the baselines optimize for hardware utilization (LLC or NUMA), they lack the awareness of interactive patterns. 
This distinctive distribution of wins proves that LAPS is not merely a static compromise, but an adaptive agent that dynamically activates the most suitable expert to boost the system performance.

\textbf{Accurate Theoretical Ceiling Approximation.}
The fundamental driver of this performance on constrained edge nodes is LAPS's accurate decision-making core. Even without any environment-specific fine-tuning, the lightweight base model correctly identifies the running workload with 97.4\% accuracy. Building on this recognition, LAPS effectively fully utilizes the limited hardware capacity by achieving 98.7\% of the theoretical performance ceiling on average (95\% CI: [97.8\%, 99.5\%]). This precision allows LAPS to systematically outmaneuver generic static optimization.

\textbf{Scalability to Multiple Concurrent Tasks.} Importantly, LAPS is fundamentally not limited to the dual-scenario combinations presented above. To explicitly verify its scalability under different concurrent tasks, we conducted additional stress tests featuring combinations of multiple co-resident applications (e.g., simultaneously running Web Browsing + Audio Remix + Network Transfer, Game Play + LLM Generate + Audio Remix, or Video Render + Audio Remix + Disk I/O). Across these multi-task scenarios, LAPS maintained a 97.9\% classification accuracy and achieved a 6.93\% average performance improvement without inducing any regressions, proving its robust capability to handle an arbitrary number of concurrent workloads.

\begin{figure}[t]
\centering
\begin{tikzpicture}
    \begin{axis}[
        axis x line*=bottom,
        axis y line*=left,
        ymajorgrids=true,
        grid style={line width=0.5pt, draw=gray!30},
        tick align=inside,
        ybar,
        bar shift=0pt,
        bar width=12pt, 
        enlarge x limits=0.08,
        width=\columnwidth, height=4cm,
        ylabel={Improvement (\%)},
        xlabel={Devices},
        symbolic x coords={VM1, VM2, VM3, VM4, VM5, VM6, VM7, VM8, VM9, VM10, BM1},
        xtick={VM1, VM2, VM3, VM4, VM5, VM6, VM7, VM8, VM9, VM10, BM1},
        xticklabel style={font=\scriptsize, align=center}, 
        ymin=0, ymax=14,
        yticklabel style={font=\scriptsize},
        nodes near coords,
        nodes near coords style={
            font=\tiny\sffamily,
            color=black,
            anchor=south,
            yshift=1pt,
            /pgf/number format/fixed,
            /pgf/number format/precision=1,
            /pgf/number format/fixed zerofill,
        },
        nodes near coords={\pgfmathprintnumber\pgfplotspointmeta\%},
        legend style={
            at={(0.99, 0.99)},
            anchor=north east,
            draw=none,
            fill=none,
            font=\scriptsize,
            cells={anchor=west}
        },
    ]
    \addplot[
        fill=figgreen,
        draw=none,
        legend image code/.code={
            \draw[fill=figgreen, draw=none] (0cm,-0.1cm) rectangle (0.3cm,0.1cm);
        }
    ] coordinates {
        (VM1, 11.7)
        (VM2, 10.1)
        (VM6, 11.2)
        (VM9, 7.0)
    };
    \addlegendentry{Prototype Machine}

    \addplot[
        fill=figblue,
        draw=none,
        legend image code/.code={
            \draw[fill=figblue, draw=none] (0cm,-0.1cm) rectangle (0.3cm,0.1cm);
        }
    ] coordinates {
        (VM3, 10.3)
        (VM4, 10.3)
        (VM5, 10.0)
        (VM7, 4.8)
        (VM8, 5.3)
        (VM10, 7.7)
        (BM1, 5.1)
    };
    \addlegendentry{Adapted Machine}

    \end{axis}
\end{tikzpicture}
\caption{Geometric mean performance improvement}
\label{fig:vm_geomean_sub}
\end{figure}

\subsection{Case Study II: Adaptability on Diverse Hardware}\label{sec:eval_claim3}
A practical intelligent system must be able to generalize from its training environment to new ones. To validate this, we compare LAPS's performance on the four prototype machines against its performance on the seven adapted ones.

As illustrated in Figure~\ref{fig:vm_geomean_sub}, LAPS delivers consistent performance gains across diverse hardware environments, validating its generalization ability. 
On the \textbf{prototype machines}, LAPS achieves a geometric mean performance improvement of \textbf{9.98\%} that of EEVDF (95\% CI: [7.83\%, 12.17\%]), while effectively converging to the theoretical limit by attaining \textbf{97.5\%} of the performance ceiling of the expert set (95\% CI: [95.85\%, 99.08\%]). 
Crucially, this efficacy translates seamlessly to the \textbf{adapted machines}. On these unseen platforms, LAPS maintains a significant lead over the baseline with a \textbf{8.03\%} average speedup (95\% CI: [6.73\%, 9.35\%]). Remarkably, its precision in policy selection remains exceptionally high, reaching \textbf{99.5\%} of the performance ceiling (95\% CI: [98.61\%, 100.37\%]). 
These tight confidence intervals confirm that LAPS's calibration pipeline captures fundamental workload characteristics rather than device-specific artifacts, ensuring robust generalization ability in real-world scenarios.

\subsection{Case Study III: Execution Safety and Fairness}\label{sec:eval_claim4}
Beyond average performance gains, a viable scheduler for embedded edge systems must guarantee robust execution and strict fairness across tasks. To confirm that LAPS provides a safe optimization path, we profile its execution safety under mispredictions and its ability to prevent background task starvation.

\textbf{Impact of Suboptimal Decisions.} 
LAPS achieves a high classification accuracy of 97.4\%. However, perfect prediction is theoretically elusive in real-world systems. We conducted a detailed analysis of the remaining 2.6\% of runtime where misrecognition occurred. 
The analysis reveals that these errors are largely benign: 68.01\% of misrecognitions still resulted in the selection of the optimal scheduler. This phenomenon stems from the fact that workloads susceptible to misclassification (e.g., distinct phases of compilation vs. encoding) often exhibit similar resource signatures, and consequently, share identical optimal scheduling policies.
Crucially, instances where LAPS selected a significantly suboptimal policy (outside the Top-3) accounted for only 0.29\% of the total runtime, averaging merely 1.04 seconds per scenario. This demonstrates that LAPS's classification errors are mostly harmless, as they rarely lead to suboptimal policy selection or long-term performance loss.

\textbf{Fairness and Starvation Avoidance.}
While the use of production-ready base schedulers mitigates basic thread starvation, a higher-level concern in interactive multitasking is whether foreground optimization penalizes background tasks. 
Specifically, we define such level of fairness as ensuring bounded progress for throughput-oriented background jobs. Neglecting this leads to starvation, where essential maintenance tasks (e.g., garbage collection, I/O flushing) stall indefinitely, destabilizing the system. To quantify this, we analyzed the performance deviation of background workloads in mixed scenarios compared to the EEVDF baseline.
Statistical analysis of background scenario scores demonstrates that LAPS maintains strict execution fairness. The average performance for background tasks improved by \textbf{+3.64\%}, indicating that resolving interactive blockers efficiently can yield positive externalities for system-wide throughput. The median performance change was negligible at \textbf{-0.01\%}. Furthermore, performance degradation in tail scenarios was strictly bounded, the 90th and 95th percentiles showed declines of only \textbf{-2.03\%} and \textbf{-3.63\%}, respectively. This demonstrates LAPS's ability to prioritize foreground responsiveness without compromising the execution integrity of background jobs.

\subsection{Ablation Studies}\label{sec:eval_ablation}

\begin{figure}[t]
\centering
\begin{tikzpicture}
  \begin{axis}[
    width=0.95\columnwidth,
    height=5cm,
    grid=both,
    minor grid style={gray!25},
    major grid style={gray!50},
    xlabel={Window length $W$ (seconds)},
    ylabel={Average response delay (s)},
    yticklabel style={/pgf/number format/fixed},
    axis y line*=left,
    axis x line*=bottom,
    legend style={at={(0.12,0.98)},anchor=north west, fill=none, draw=none},
    xmin=0, xmax=20,
  ]
    \addplot+[blue, mark=*, thick]
      table[x=window_sec, y=avg_delay_sec, col sep=comma]
      {figures/window_length_sensitivity.csv};
    \addlegendentry{Avg delay (s)}
  \end{axis}

  \begin{axis}[
    width=0.95\columnwidth,
    height=5cm,
    axis y line*=right,
    axis x line=none,
    ylabel={Error rate (\%)},
    yticklabel style={/pgf/number format/fixed},
    legend style={at={(0.98,0.98)},anchor=north east, fill=none, draw=none},
    xmin=0, xmax=20,
    ymin=3, ymax=5.5,
  ]
    \addplot+[red, mark=square*, thick]
      table[x=window_sec, y=error_rate_percent, col sep=comma]
      {figures/window_length_sensitivity.csv};
    \addlegendentry{Error rate (\%)}
  \end{axis}
\end{tikzpicture}
\caption{Sensitivity of Time-Weighted Probability Voting to window length}
\label{fig:window_length_sensitivity}
\end{figure}

To isolate the impact of key design choices, we conducted ablation studies.

\textbf{Impact of Window Length.}
A critical component of LAPS is the Time-Weighted Probability Voting mechanism, designed to ensure decision stability by filtering out transient system noise. The window size, denoted as $W$, defines the temporal scope of the historical workload probability predictions. Specifically, at any decision timestamp $t$, the Time-Weighted Probability Voting mechanism aggregates the historical probability distributions over the interval $(t-W, t]$ to compute the weighted consensus. Figure~\ref{fig:window_length_sensitivity} shows its impact by varying the window length $W$.
The results show a clear U-shaped curve for response delay. 
A very short window ($W<4s$) leads to high average delay because the system becomes overly sensitive to transient noise, causing frequent, erroneous scheduler switches. This policy ``thrashing'' means the system spends significant time in a suboptimal state while oscillating, which paradoxically increases the time required to settle on the correct policy. 
Conversely, a very long window ($W>10s$) also increases delay by making the system too slow to respond to genuine workload changes. The plot reveals a Pareto-optimal range between 4s to 7s. We chose $W=6s$ as a robust compromise. Although this still introduces a limitation in rare high-frequency switching scenarios, as the triggered physical cooldown causes the system to temporarily utilize a suboptimal scheduler, this configuration overall provides the most balanced responsiveness and stability. It enables LAPS to make confident decisions and achieve reliable performance optimization.

\begin{figure}[t]
  \begin{tikzpicture}
    \begin{axis}[
        trim axis left, 
        trim axis right,
        width=\columnwidth,
        height=5cm,
        tick align=inside,
        xlabel={Scheduler Set Size},
        xlabel style={yshift=-3pt},
        xmin=0.5, xmax=8.5,
        xtick={1,2,3,4,5,6,7,8},
        ylabel={Geo Mean Improvement (\%)},
        ymin=-0.5, ymax=11.5,
        ytick={0,2,4,6,8,10},
        grid=major,
        grid style={line width=0.5pt, color=gray!20},
        extra y ticks={0},
        extra y tick style={grid=major, grid style={dashed, gray!80}},
        clip=false,
        axis x line*=bottom,
        axis y line*=left,
    ]

    \addplot[
        color=figblue,
        mark=*,
        mark options={fill=figblue, scale=1.2},
        line width=1.5pt,
        sharp plot
    ] coordinates {
        (1, 0.00) (2, 5.64) (3, 7.42) (4, 8.40)
        (5, 8.96) (6, 9.36) (7, 9.53) (8, 9.62)
    };

    \tikzstyle{valbox} = [draw=gray!60, fill=white, rounded corners=2pt, inner sep=1.5pt, anchor=south, yshift=4pt, font=\scriptsize]
    
    \tikzstyle{namebox} = [fill=yellow!20, inner sep=1.5pt, anchor=north, yshift=-2.8pt, font=\scriptsize]
    
    \tikzstyle{namebox_low} = [fill=yellow!20, inner sep=1.5pt, anchor=north, yshift=-10pt, font=\scriptsize]
    \tikzstyle{namebox_low2} = [fill=yellow!20, inner sep=1.5pt, anchor=north, yshift=-5pt, font=\scriptsize]
    \tikzstyle{namebox_low3} = [fill=yellow!20, inner sep=1.5pt, anchor=north, yshift=-7pt, font=\scriptsize]
    \tikzstyle{namebox_low6} = [fill=yellow!20, inner sep=1.5pt, anchor=north, yshift=-6pt, font=\scriptsize]
    \tikzstyle{namebox_low9} = [fill=yellow!20, inner sep=1.5pt, anchor=north, yshift=-2.6pt, font=\scriptsize]

    \node[valbox] at (axis cs:1, 0.00) {0.00\%};
    
    \node[valbox] at (axis cs:2, 5.64) {5.64\%};
    \node[namebox] at (axis cs:2, 5.64) {scx\_flash};

    \node[valbox] at (axis cs:3, 7.42) {7.42\%};
    \node[namebox_low3] at (axis cs:3, 7.42) {scx\_lavd};

    \node[valbox] at (axis cs:4, 8.40) {8.40\%};
    \node[namebox_low3] at (axis cs:4, 8.40) {scx\_simple};

    \node[valbox] at (axis cs:5, 8.96) {8.96\%};
    \node[namebox_low2] at (axis cs:5, 8.96) {scx\_bpfland};

    \node[valbox] at (axis cs:7, 9.53) {9.53\%};
    \node[namebox_low] at (axis cs:7, 9.53) {scx\_p2dq};
    
    \node[valbox] at (axis cs:6, 9.36) {9.36\%};
    \node[namebox_low9] at (axis cs:6, 9.36) {scx\_rusty};

    \node[valbox] at (axis cs:8, 9.62) {9.62\%};
    \node[namebox] at (axis cs:8, 9.62) {scx\_nest};

    \end{axis}
\end{tikzpicture}
  \caption{Cumulative Performance Improvement vs. Scheduler Set Size. The curve flattens as the set expands, indicating diminishing marginal benefit. However, the diverse set is essential to cover the long tail of workload patterns effectively.}
  \label{fig:scheduler_set_ablation}
\end{figure}

\textbf{Impact of Scheduler Set Size.}
Another key question is about the impact of the size of expert scheduler set. Figure~\ref{fig:scheduler_set_ablation} illustrates the cumulative geometric mean performance improvement relative to EEVDF as we incrementally add expert schedulers to the set.%
The results show a characteristic diminishing returns curve. The first few schedulers provide the steepest initial gains. As the set size grows beyond 4-5 schedulers, the marginal improvement per additional scheduler decreases.%
Critically, however, the curve's flattening tail does not imply that the later additions are redundant. While their average contribution across \textit{all} scenarios is smaller, they are often the definitive optimal choice for specific, boundary-case load patterns where generic policies fail. For example, \emph{scx\_nest} might only offer significant gains in a narrow slice of scenarios, but in those slices, they outperform others by a wide margin. 
Therefore, maintaining a larger, more diverse scheduler set allows LAPS to perform fine-grained matching across a wider variety of device configurations and workload scenarios, ensuring robustness even in edge cases. This validates our design choice of supporting an extensible, broad scheduler portfolio rather than relying on a small fixed set.

\subsection{System Overhead}\label{sec:eval_overhead}
To verify our solution to \textbf{Challenge 2} (Confining Intelligence within Strict Embedded Resource Budgets), we evaluate LAPS's resource consumption and latency across all test scenarios. On average, LAPS's user-space agent and kernel-level components collectively consume just 0.59\% of a single CPU core and 24.83 MB of memory. This level of overhead is negligible even for edge computing nodes.

\begin{table}[htbp]
\centering
\caption{LAPS Latency Statistics (values in milliseconds)}
\label{tab:agent_latency}
\begin{tabular}{l|r|r|r|r}
\hline
\textbf{Latency Type} & \textbf{Average} & \textbf{$P_{95}$} & \textbf{$P_{99}$} & \textbf{Max} \\
\hline
Inference Latency & 0.251  & 0.308 & 0.320 & 0.322\\
Decision Latency & 0.038  & 0.050 & 0.053 & 0.054\\
Total Latency & 0.289  & 0.357 & 0.367 & 0.369\\
\hline
\end{tabular}
\end{table}

To assess the responsiveness of LAPS, we measured the latency of its decision-making pipeline, breaking it down into two main components. \textit{Inference latency} covers the time from data collection to the prediction of the workload's probability distribution. \textit{Decision latency} is the time from the completion of inference to the determination of a scheduling policy switch. The total latency is the sum of these two.
Table~\ref{tab:agent_latency} presents the average and percentile latencies for these components. To understand worst-case behavior, we examine the 95th, and 99th percentile ($P_{95}$, $P_{99}$) latencies. These values indicate that 95\%, and 99\% of the decisions, respectively, are made faster than these thresholds. The average total latency is a mere 0.289 ms, while the max latency is only 0.369 ms. As LAPS's Observe-Infer-Switch cycle runs on a \emph{seconds-level} periodic basis, its \emph{microseconds-level} latency confirms that LAPS can react to system state changes in a timely manner, even under heavy load. In summary, LAPS's benefits are delivered at a minimal and acceptable cost.

\subsection{Summary}\label{sec:eval_summary}
Our comprehensive evaluation demonstrates that LAPS successfully validates its core premise: dynamic, intelligent selection of expert schedulers is a superior paradigm for modern operating systems. The experiments confirm that LAPS significantly outperforms a mainstream general-purpose scheduler (EEVDF), effectively approaches the performance ceiling of our scheduler set, and generalizes robustly to new hardware environments. Ablation studies verify that its decision stability mechanisms are critical to its success, and overhead analysis confirms its real-world practicality.

\section{Conclusion}
By designing and implementing the Lightweight Adaptive Policy Switcher, this study has verified the feasibility and effectiveness of the ``dynamically switching between expert policies'' design philosophy in the field of operating system scheduling. LAPS not only achieves technical innovation but, more importantly, provides new ideas and methods for development of intelligent operating system schedulers.

As edge computing environments continue to evolve and application workloads become increasingly complex, traditional static scheduling policies struggle to meet the performance demands of modern systems. The adaptive, intelligent scheduling direction represented by LAPS provides a feasible technical path to address this challenge, and the results of this research not only offer substantial academic value but also provide practical insights for industrial applications.

\appendices
\section{Model Hyperparameters}

To ensure reproducibility, Table~\ref{tab:hyperparameters} details the optimal hyperparameters used for training the XGBoost classifier in the Inference module. We utilized the Optuna framework to conduct hyperparameter tuning experiments and obtain these final parameter configurations.

\begin{table}[htbp]
  \centering
  \caption{Optimal Hyperparameters for XGBoost Classifier}
  \label{tab:hyperparameters}
  \renewcommand{\arraystretch}{1.2}
  \begin{tabular}{l c | l c}
    \toprule
    \textbf{Hyperparameter} & \textbf{Value} & \textbf{Hyperparameter} & \textbf{Value} \\
    \midrule
    \texttt{n\_estimators} & 2000 & \texttt{max\_depth} & 9 \\
    \texttt{colsample\_bytree} & 0.52 & \texttt{subsample} & 0.80 \\
    \texttt{learning\_rate} & 0.22 & \texttt{gamma} & 0.12 \\
    \texttt{min\_child\_weight} & 3.24 & \texttt{reg\_alpha} & 0.06 \\
    \texttt{early\_stopping\_rounds} & 50 & \texttt{reg\_lambda} & 8.30 \\
    \bottomrule
  \end{tabular}
\end{table}

\section{Detailed Benchmark Scenarios Design}

This appendix provides detailed construction methodologies and tools used for the 11 benchmark scenarios evaluated in Section V of the main manuscript. These workloads are specifically designed to simulate the complex resource contention patterns found in modern smart vehicle cabins and edge computing environments.

\textbf{Web Browsing}: Emulates interactive web-based UI applications in smart vehicle cabins requiring immediate responsiveness. We used Selenium to automate interactions (clicks, scrolls, tab switches) in Chrome on local HTML files. JavaScript timers recorded response times and frame rate stability.

\textbf{Game Play}: Emulates in-car entertainment systems requiring stable frame generation. We used Minecraft with a Fabric API component to simulate continuous player movement on a fixed map, recording frame rate, frame generation time, loaded chunks, and milliseconds per game tick (MSPT).

\textbf{Office File Editing}: Emulates interactive applications and document processing in edge environments requiring immediate responsiveness. We utilized LibreOffice and the UNO framework to automate text editing, formatting, and cross-document copying, measuring input and rendering latencies.

\textbf{Audio Remix}: Models the concurrent playback and mixing of media and navigation prompts in smart vehicles, representing latency-sensitive soft real-time tasks. PulseAudio was used to perform mixing and apply effects, recording buffer underruns (XRuns) and computation latency.

\textbf{Disk I/O}: Reflects bursty data transmission demands and embedded I/O bottlenecks. The test involves creating a mix of large, medium, and small files, and executing a combination of sequential and random read/write operations via multiple processes, recording IOPS and data transfer rates.

\textbf{LLM Generate}: Represents edge AI inference applications, such as in-car voice assistants, that impose computational and memory access loads. We evaluate LLM inference using llama.cpp with the DeepSeek-R1-Distill-Qwen-1.5B model \cite{deepseekai2025deepseekr1}. The test simulates continuing a conversation from a 400-token context and generating approximately 1000 tokens per request, recording time-to-first-token and tokens-per-second.

\textbf{Network Transfer}: Reflects data transmission demands and dense networking events in edge environments. iperf3 was used to generate high-throughput TCP and UDP data transmission, recording transmission rates and packet loss percentages.

\textbf{Video Render}: Emulates the computational requirements of multi-channel camera video stream encoding in automotive systems. FFmpeg was used to perform video encoding and transcoding tasks, measuring total completion time and average rendering rate.

\textbf{Blender Render}: Emulates the computational requirements of 3D digital twins and surround-view stitching in vehicle interfaces. Blender was used to render a 3D scene using available CPU cores, recording tile rendering speed and total rendering time.

\textbf{Kernel Compile}: Reflects the computation needs and file system access inherent to continuous Over-The-Air (OTA) updates in vehicle systems. The test involves compiling the Linux 6.14 kernel under default configuration and recording the total compilation time.

\textbf{Archive Extraction}: Reflects the computation needs and file system access inherent to OTA updates and asset installations in vehicle systems. 7-Zip was used to perform compression and decompression tasks, measuring the processing rate.

\bibliographystyle{IEEEtran}
\bibliography{0_reference}

\begin{IEEEbiography}[{\includegraphics[width=1in,height=1.25in,clip,keepaspectratio]{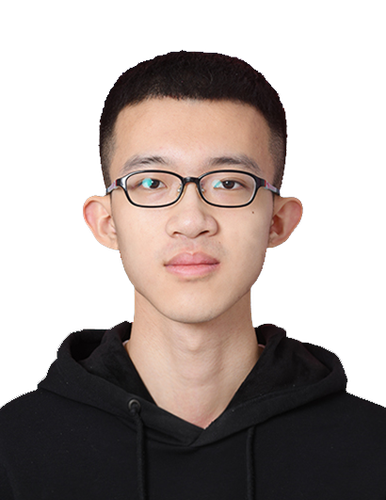}}]{Xinbo Wang}
received the B.S. degree from Zhejiang University in 2025. He is currently a graduate student at Zhejiang University. His research interests include agentic operating system, general kernel extensibility, and system efficiency optimization.
\end{IEEEbiography}

\begin{IEEEbiography}[{\includegraphics[width=1in,height=1.25in,clip,keepaspectratio]{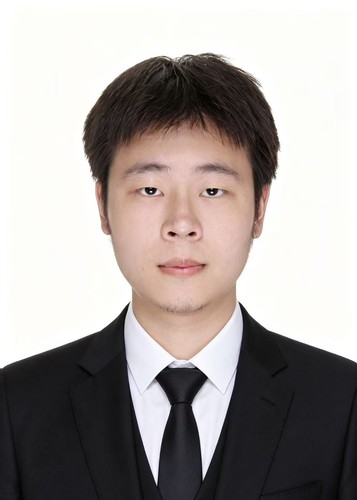}}]{Shian Jia}
received the B.S. degree from Zhejiang University in 2024. He is currently a graduate student at Zhejiang University. His research interests include agentic operating systems, agent memory architectures, and knowledge representation.
\end{IEEEbiography}

\begin{IEEEbiography}[{\includegraphics[width=1in,height=1.25in,clip,keepaspectratio]{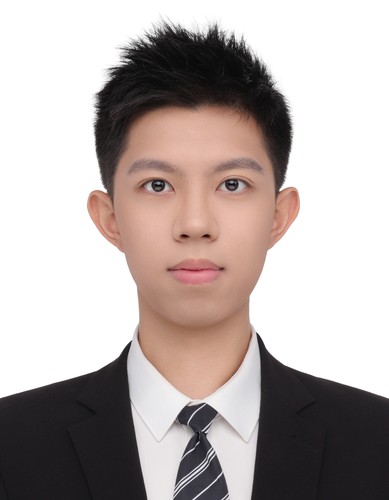}}]{Ziyang Huang}
received the B.S. degree from Zhejiang University in 2026. He is currently pursuing an M.S. degree at the same institution. His research interests primarily focus on agentic operating system, and kernel programmability and extensibility.
\end{IEEEbiography}

\begin{IEEEbiography}[{\includegraphics[width=1in,height=1.25in,clip,keepaspectratio]{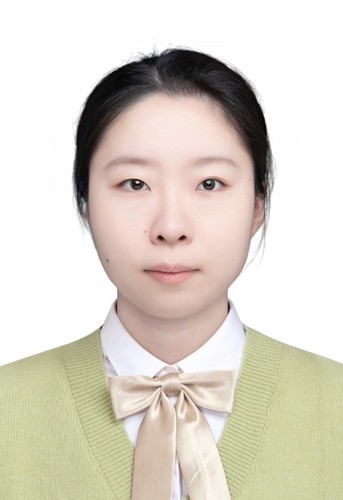}}]{Jing Cao}
received the B.S. degree from Zhejiang University in 2018 and the Ph.D. degree from the University of Science and Technology of China in 2024. She is currently an Assistant Research Fellow in Hangzhou City University. Her research interests include model training, agentic operating system and system efficiency optimization.
\end{IEEEbiography}

\begin{IEEEbiography}[{\includegraphics[width=1in,height=1.25in,clip,keepaspectratio]{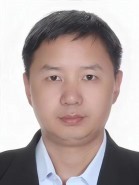}}]{Mingli Song}
(Senior Member, IEEE) received the PhD degree in computer science from Zhejiang University, China, in 2006. He is currently a professor with the institute of computer software, Zhejiang University. His research interests include face modeling and facial expression analysis. He received the Microsoft Research Fellowship in 2004.
\end{IEEEbiography}

\end{document}